\documentclass[preprintnumbers,nofootinbib,amsmath,amssymb,twocolumn]{revtex4}
\pdfoutput=1
\usepackage{amsmath}
\usepackage{amsfonts}
\usepackage{amssymb}
\usepackage{graphicx}    
\usepackage{amsfonts}
\usepackage{amsmath}
\usepackage{braket}
\usepackage{rotating}
\usepackage{verbatim}
\usepackage{multirow}

\newcommand{\fnl}{$f_{\rm NL}$}
\newcommand{\bicep}{{\sc Bicep2}}
\newcommand{\planck}{{\sc Planck}}

\begin{document}

\preprint{Imperial/TP/2014/CC/3} \vskip 0.2in

\title{BICEP's bispectrum}

\author{Jonathan~S.~Horner}
\affiliation{Theoretical Physics, Blackett Laboratory, Imperial College, London, SW7 2BZ, UK}
\author{Carlo~R.~Contaldi}
\affiliation{Theoretical Physics, Blackett Laboratory, Imperial College, London, SW7 2BZ, UK}
\affiliation{Canadian Institute of Theoretical Physics, 60 St. George
  Street, Toronto, M5S 3H8, On, Canada}
\date{\today}

\begin{abstract}
The simplest interpretation of the \bicep\ result is that the scalar
primordial power spectrum is slightly suppressed at large scales
\cite{Contaldi:2014zua,Contaldi:2014rna}. These models result in a large tensor-to-scalar
ratio $r$. In this work we show that the type of inflationary
trajectory favoured by \bicep\ also leads to a larger non-Gaussian
signal at large scales, roughly an order of magnitude larger than a
standard slow-roll trajectory. 
\end{abstract}

\maketitle

\section{Introduction}

The recent results from \bicep\ \cite{Ade:2014xna}, hinting at a
detection of primordial $B$-mode power in the Cosmic Microwave
Background (CMB) polarisation, place the inflationary paradigm on much
firmer footing. This result, in combination with the \planck\ total
intensity measurement \cite{Ade:2013ktc}, imply that primordial
perturbations are generated from an almost de-Sitter like phase of
expansion early in the Universe's history before the standard big bang
scenario.

At first glance there is potential tension between the polarisation
measurements made by \bicep\ and \planck's total intensity
measurements. \planck's power spectrum is lower than the best-fit
$\Lambda$CDM models at multipoles $\ell \lesssim 40$ and \bicep's high
$B$-mode measurement exacerbates this since tensor modes also
contribute to the total intensity. The tension is indicated by the
difference in the $r\sim 0.2$ value implied by \bicep's measurements
and the 95\% limit of $r < 0.1$ implied by the \planck\ data for
$\Lambda$CDM models. Many authors have pointed out how the tension can
be alleviated by going beyond the primordial power-law, $\Lambda$CDM
paradigm by allowing running of the spectral indices, enhanced
neutrino contributions (see for examples \cite{Czerny:2014wua,Contaldi:2014zua,Gong:2014cqa,Zhang:2014dxk}) or
more exotic scenarios \cite{Anchordoqui:2014dpa}. However the simplest explanation,
that also fits the data best, is one where there is a slight change in
acceleration trajectory during the inflationary phase when the largest
modes were exiting the horizon. This was shown by \cite{Contaldi:2014zua} where
a specific model was used to generate a slightly faster rolling
trajectory at early times. The effect of such a ``slow-to-slow-roll''
transition is to result in a slightly suppressed primordial, scalar
power spectrum that fits the \planck\ data despite the large tensor
contribution required by \bicep. In \cite{Contaldi:2014rna} the author analyses
generalised accelerating, or inflating, trajectories that fit the
combination of \bicep\ and \planck\ data and conclude that the
suppression is required at a significant level and the best-fit
trajectories are all of the form where the acceleration has a slight
enhancement at early times.

An alternative explanation is that the $B$-mode power observed by
\bicep\ is not due to foregrounds and is not primordial. This
possibility has been discussed by various authors
\cite{Mortonson:2014bja,Flauger:2014qra} who point out that more measurements on the
frequency dependence of the signal are required to definitively state
whether we have detected the signature of primordial tensor
modes. These measurements will be provided in part by the \planck\
polarisation analysis and \bicep's cross-correlation with further KECK
data \cite{Ade:2014xna}. 

If the \bicep\ result stands the test of time then the signal we point
out in the analysis below is expected to be present if the simplest
models of inflation driven by a single, slow-rolling scalar field are
the explanation behind the measurements. In this case a measurement of
tensor mode amplitude, or $r$, is a direct measurement of the
background acceleration since $r \sim 16\epsilon$ and the tension
between \bicep\ polarisation and \planck\ total intensity measurements
implies a change in the acceleration at early times. In turn, the
change in acceleration enhances the non-Gaussianity on scales that
were exiting the horizon while the acceleration was changing. 

In this {\sl paper} we construct a simple toy-model inspired by the best
fitting trajectories found in \cite{Contaldi:2014rna} and calculate its
bispectrum numerically. At small scales, as one would expect, the
non-Gaussianity is small $\mathcal{O}(10^{-2})$ \cite{Maldacena:2002vr,Creminelli:2004yq} but at large scales,
where the scalar power spectrum is suppressed, the non-Gaussianity can
be significantly larger, $\mathcal{O}(10^{-1})$. The results are compared
against the slow-roll approximation in the equilateral configuration
and the squeezed limit consistency relation. Whilst at small scales
there is exceptional agreement with the slow-roll approximation, at
large scales the results can deviate by up to 10\%.

This {\sl paper} is organised as follows. We outline the calculation of the
scalar and tensor power spectra in Section \ref{power_spec} and
summarise the calculation of the bispectrum in Section
\ref{bispectrum}. Our results are presented in Section \ref{results}
and we discuss their implications in Section \ref{conclusion}.

\section{Computation of the scalar power spectrum}\label{power_spec}

The calculation is best performed in a gauge where all the scalar
perturbations are absorbed into the metric such that $g_{ij} =
a^{2}\,(t)e^{2\zeta(t, \mathbf{x})}\delta_{ij}$ and the inflaton
perturbation $\delta\phi(t, \mathbf{x}) = 0$. The primordial power
spectrum is then simply given by:
\begin{equation}\label{PowerSpectrum}
\langle\zeta^{}_{k_{1}}\zeta^\star_{k_{2}}\rangle = (2\pi)^{3}\delta^{(3)}(\mathbf{k}_{1}+\mathbf{k}_{2})P_\zeta(k_1)\,,
\end{equation}
where $\mathbf{k}$ is the Fourier wavevector and $k\equiv |\mathbf{k}|$. The mode $\zeta_{k}(t)$ satisfies the Mukhanov-Sasaki equation \cite{Mukhanov:1985rz,Sasaki:1986hm}
\begin{equation}\label{Mukh}
  \frac{\mathrm{d}^{2}\zeta_{k}}{\mathrm{d}N^{2}} + (3 + \epsilon - 2\eta)\frac{\mathrm{d}\zeta_{k}}{\mathrm{d}N} + \frac{k^{2}}{a^{2}H^{2}}\zeta_{k} = 0\,.
\end{equation} 
In the above $N$ is the number of $e-$folds which increases with time or alternatively
\begin{equation}
H = \frac{\dot{a}}{a} = \frac{dN}{dt}\,,
\end{equation}
and $\epsilon$ and $\eta$ are the usual slow-roll variables defined by
\begin{equation}
\epsilon = -\frac{\dot{H}}{H^{2}},\,\,\,\,\,\,\,\, \eta = \epsilon - \frac{1}{2H}\frac{d\ln\epsilon}{dt}\,.
\end{equation}

Outside the horizon $\zeta_{k}$ quickly goes to a constant and the
power spectrum is then related to the freeze-out value of $\zeta_k$ on
scales $k\ll aH$
\begin{equation}
  P_\zeta(k) = \left|\zeta_{k\ll aH}\right|^{2}\,.
\end{equation}
The initial conditions for the solutions to (\ref{Mukh}) can be set
when the mode is much smaller than the horizon $k\gg aH$ and takes on
the  Bunch-Davies  form \cite{Bunch:1978yq}
\begin{equation}\label{Initial_zeta}
\zeta_{k} \to \frac{1}{M_{\rm pl}}\,\frac{e^{-ik\tau}}{2a\sqrt{k\epsilon}}\,,
\end{equation}
where $\tau$ is conformal time defined by $\mathrm{d}N/\mathrm{d}\tau =
aH$. 

An identical calculation can be performed for the tensor power
spectrum $P_{h}(k) = \left|h_{k\ll aH}\right|^{2}$ with $h_{k}$
satisfying the following differential
equation 
\begin{equation}\label{tensors}
  \frac{\mathrm{d}^{2}h_{k}}{\mathrm{d}N^{2}} + (3 -
  \epsilon)\frac{\mathrm{d}h_{k}}{\mathrm{d}N} +
  \frac{k^{2}}{a^{2}H^{2}}h_{k} = 0\,,
\end{equation} 
with initial condition
\begin{equation}
h_{k} \to \frac{1}{M_{\rm pl}}\,\frac{e^{-ik\tau}}{a\sqrt{2k}}\,,
\end{equation}
in the limit where $k \gg aH$. Solving for $P_{\zeta}(k)$ and $P_{h}(k)$ numerically we can calculate $n_{s}, r$ and $n_{t}$ directly from their definitions:
\begin{eqnarray}\label{nsr_numeric}
n_{s}(k_\star) & = & 1 + \left.\frac{\mathrm{d}\ln \left[k^{3}P_{\zeta}(k)\right]}{\mathrm{d}\ln k}\right|_{k = k_\star}\,\\
r(k_\star) & = & 8\,\frac{P_h(k_\star)}{P_{\zeta}(k_\star)}\,\nonumber\\\nonumber
n_{t}(k_\star) & = & \left.\frac{\mathrm{d}\ln \left[k^{3}P_{h}(k)\right]}{\mathrm{d}\ln k}\right|_{k = k_\star}\,\\\nonumber
\end{eqnarray}
The factor of 8 comes from how the tensor perturbations are normalised in the second order action.

The above procedure outlines the general calculation of the primordial
power spectrum from inflation. In this work we are interested in
specifying a background model favoured by the recent \bicep\ + \planck\
data. In particular we choose a function for $\epsilon$, then $\eta$
and $H$ are easily obtained by its derivative and integral
respectively.

\begin{figure}[t]
  \begin{center}
    \includegraphics[width=8.5cm,trim=0cm 0cm 0cm 0cm,clip]{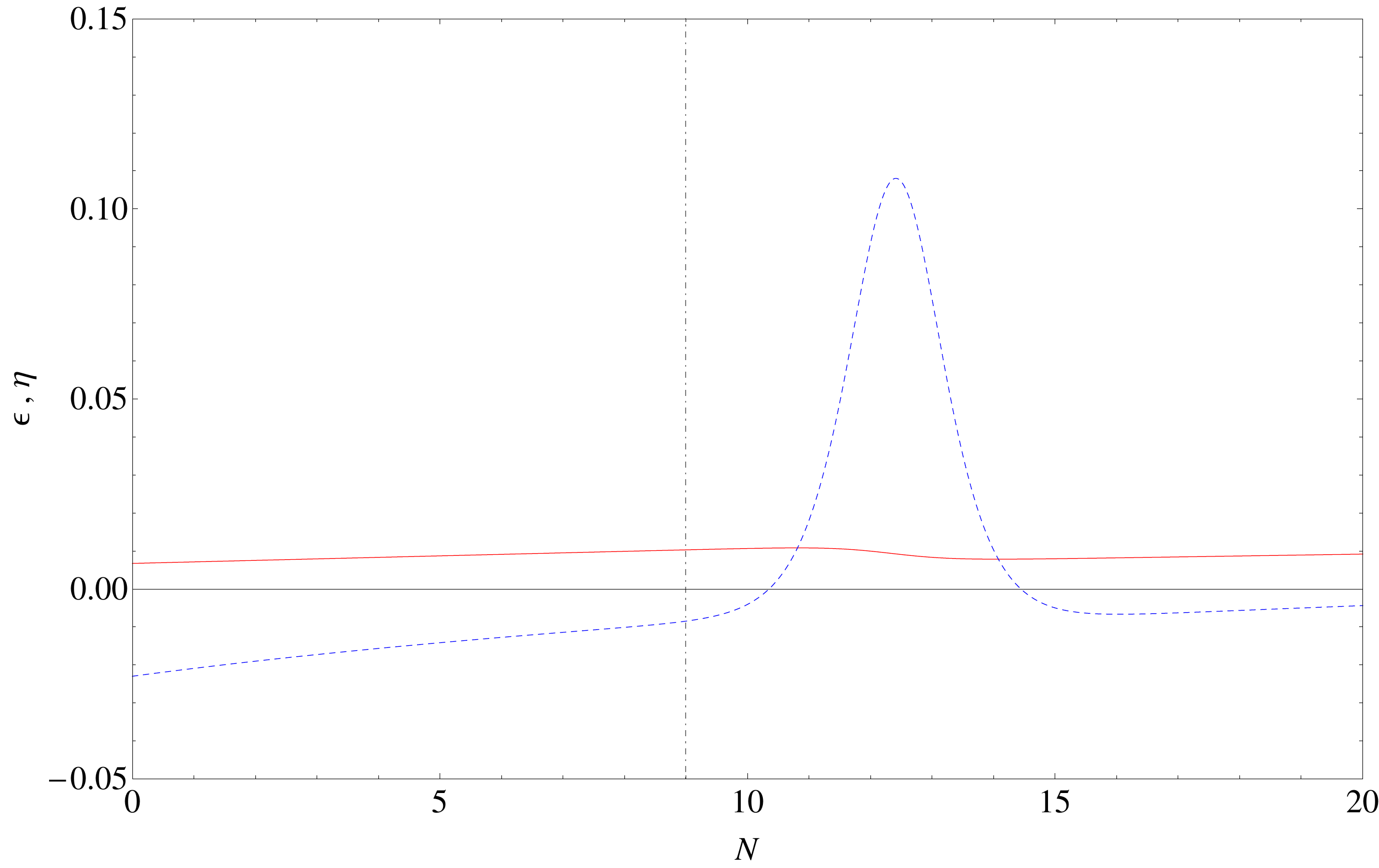} 
    \caption{ Background functions $\epsilon$ (red, solid) and $\eta$ (blue, dashed) of our toy-model plotted as a function of $e-$folds $N$. The grey vertical line indicates roughly the time when the first observable mode crosses the horizon.}
    \label{fig:background}
  \end{center}
\end{figure}

\begin{figure*}[t]
  \begin{center}
    \begin{tabular}{cc}
      \makebox[8.5cm][c]{
        \includegraphics[width=8.5cm,trim=0cm 0cm 0cm
        0cm,clip,angle=0]{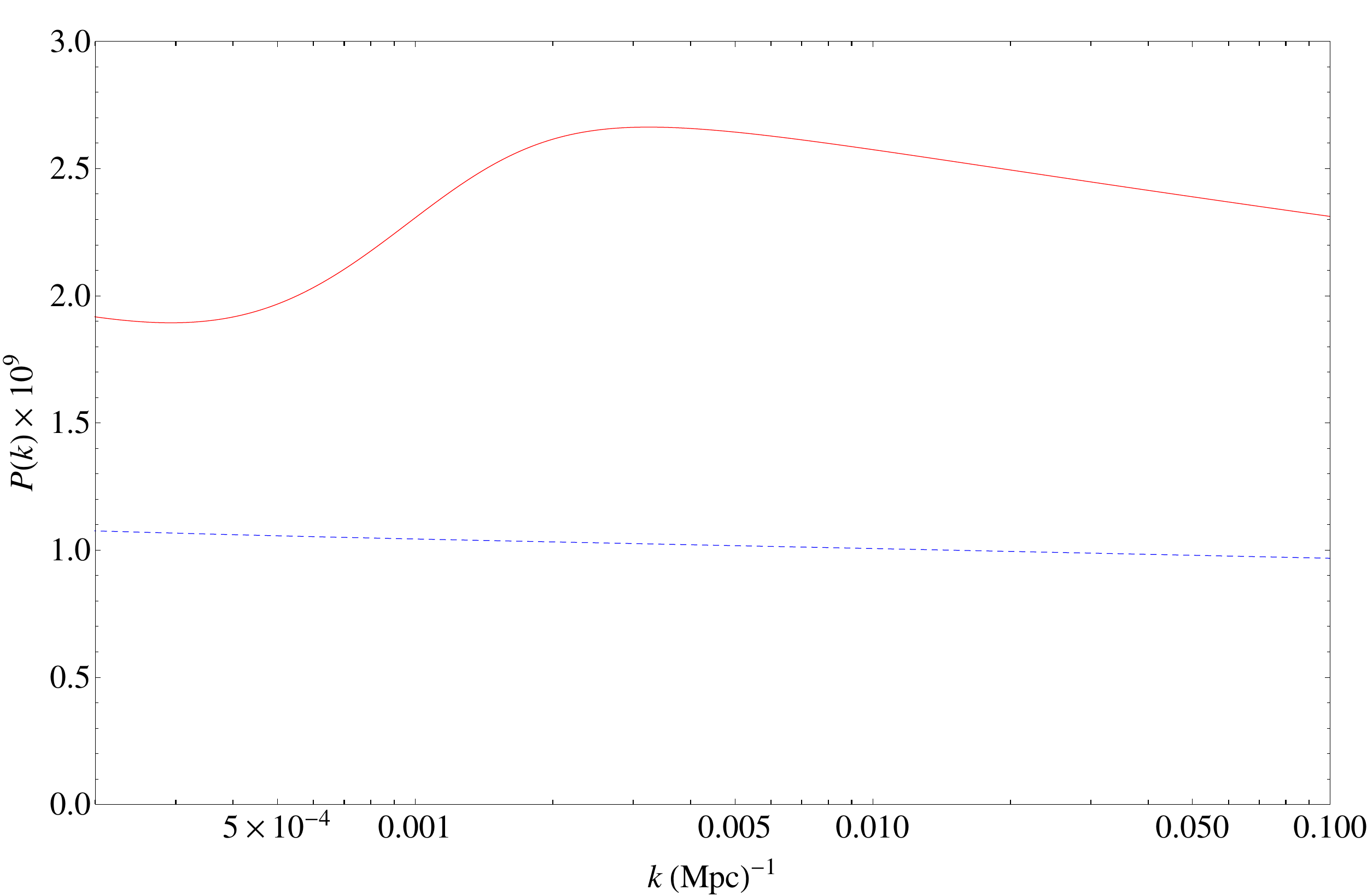}}&
      \makebox[8.5cm][c]{\includegraphics[width=8.5cm,trim=0cm 0cm 0cm
        0cm,clip,angle=0]{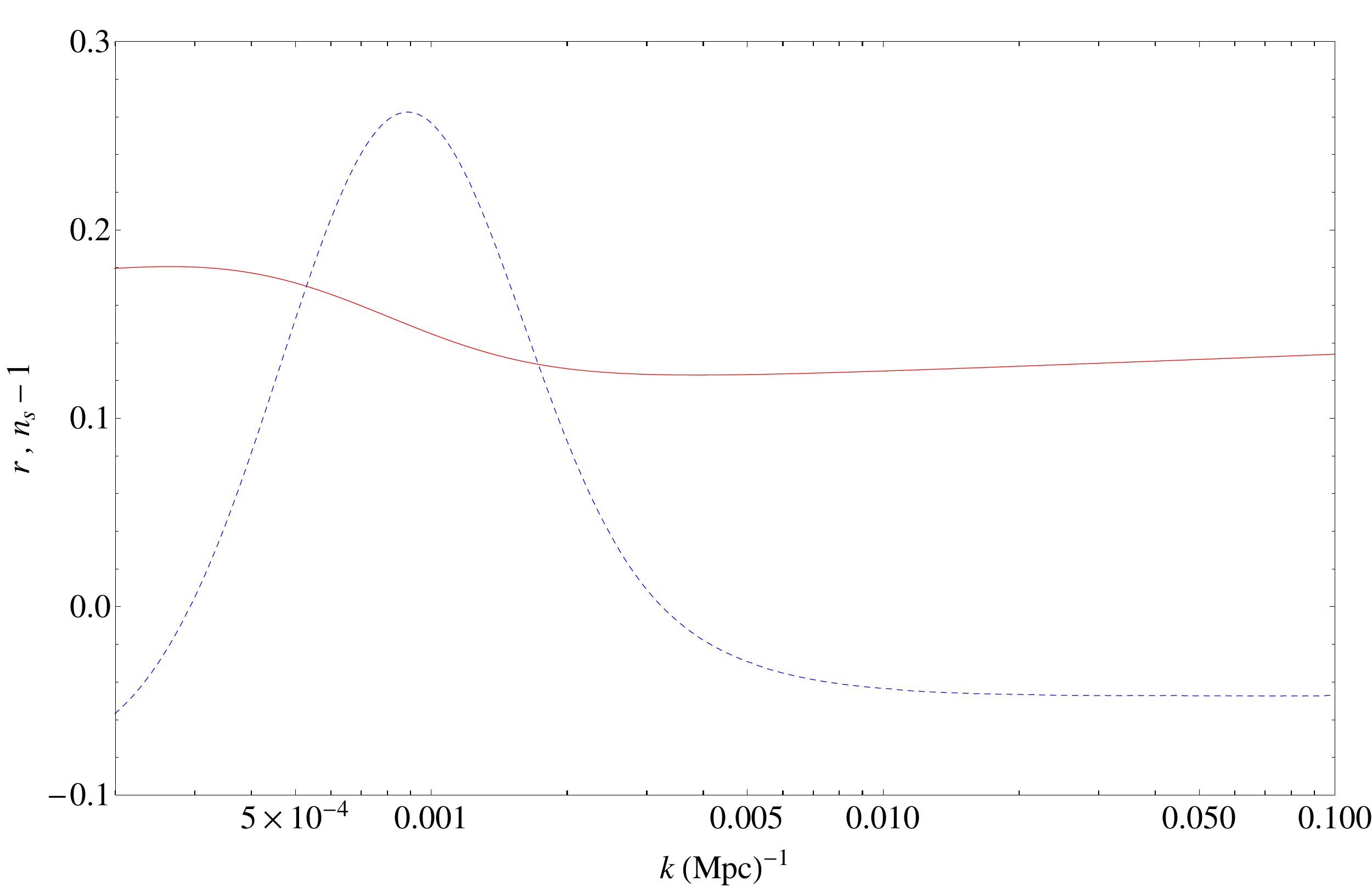}}
    \end{tabular}
    \caption{Left: Scalar (red, solid) and tensor (blue, dashed)
      dimensionless power-spectra. The tensors have been multiplied by
      a factor of 25 for comparison. Right: $r$ (red, solid) and
      $n_{s} - 1$ as functions of $k$. The parameters in the toy-model
      were chosen to give a good match to the \planck\ and
      \bicep\ data.}
    \label{fig:ps}
  \end{center}
\end{figure*}

Instead of a direct function of time or $N$ though we specify
$\epsilon(x)$ where $x = \ln (k'/k_{\rm min})$. $k'$ is the mode
crossing the horizon at $e-$foldings $N$ ($k' = aH$) and $k_{\rm min} \sim
10^{-5} (\text{Mpc})^{-1}$ is the largest scale observable today. In
addition to being proportional to $r$ this condition allows one to
easily specify how the background should evolve in our observational
window. For concreteness we require $\epsilon$ to be relatively large,
but still satisfying the slow-roll limit, 
at large scales and then to flatten out into another slow-roll regime
with a smaller value. To this end we adopt a simple toy-model for
$\epsilon$ as a function of x
\begin{equation}\label{toy_model}
\epsilon = \left\{\epsilon_1 \tanh\left[(x - x_0)\right] + \epsilon_2\right\}\left(1 + m x\right)\,,
\end{equation}
where the coefficients $\epsilon_1$, $\epsilon_2$,  $m$, and
$x_0$ are chosen to give a final power spectrum with the required
suppression and position ($\sim 26\%$ and 1.5$\times 10^{-3}$
Mpc$^{-1}$ respectively \cite{Contaldi:2014zua}) and $n_s\sim 0.96$ on small
scales.  Fig.~\ref{fig:background} shows $\epsilon$ and $\eta$ as a
function of $N$ for this toy-model and the resulting power spectra are
shown in Fig.~\ref{fig:ps}.

\section{Computation of the bispectrum}\label{bispectrum}

The largest contribution to primordial non-Gaussianity will come from the bispectrum of the curvature perturbation
\begin{equation}\label{3rd}
\langle\zeta_{k_{1}}\zeta_{k_{2}}\zeta_{k_{3}}\rangle  =  (2\pi)^{3}\delta^{(3)}(\mathbf{k}_{1}+\mathbf{k}_{2}+\mathbf{k}_{3})B(k_{1}, k_{2}, k_{3})\,.
\end{equation}
The quantity that is often quoted in observational constraints is the dimensionless, reduced bispectrum
\begin{eqnarray}\label{fnl_def}
  f_{\mathrm{NL}}(k_{1}, k_{2}, k_{3})  &=&  \frac{5}{6}\,B(k_{1}, k_{2},
  k_{3}) / \left(|\zeta_{k_{1}}|^{2}|\zeta_{k_{2}}|^{2}+\right.\nonumber\\
  &&\left.|\zeta_{k_{1}}|^{2}|\zeta_{k_{3}}|^{2}+|\zeta_{k_{2}}|^{2}|\zeta_{k_{3}}|^{2}\right)\,,
\end{eqnarray}
The analytical calculation is much simpler if we consider the
equilateral configuration $f_{\text{NL}}(k,k,k)$ however this is not a
directly observed quantity as the estimator requires $B(k_{1},
k_{2}, k_{3})$ to be factorizable \cite{Creminelli:2005hu}. This is not true for the general
case, which we are considering. However the overall amplitude of the
reduced bispectrum gives a good indication of the size of the expected
observable $f_{\mathrm{NL}}$.
 
All theories of inflation will produce a non-zero bispectrum. This is
simply because gravity coupled to a scalar field is a non-linear
theory and will contain interaction terms for the primordial curvature
perturbation $\zeta(t,\textbf{x})$. These interaction terms will
source the bispectrum with the largest contributors coming from
tree-level diagrams associated with the cubic interaction terms. The
bispectrum can then be calculated using the ``in-in'' formalism
\cite{Adshead:2009cb,Maldacena:2002vr,Seery:2005gb}, which to tree
level becomes
\begin{equation}\label{in_in}
\langle\zeta^{3}(t)\rangle = -i\int_{-\infty}^{t}\mathrm{d}t'\langle\left[\zeta^{3}(t), H_{\text{int}}(t')\right]\rangle\,,
\end{equation}
where $H_{\text{int}}$ is the interaction Hamiltonian associated with the following third order action
\begin{eqnarray}\label{action_final}
  S_{3} &=& \!\!\int d^4x\, a^{3}\epsilon \left[\left(2\eta - \epsilon\right)\zeta \dot{\zeta}^{2} + \frac{1}{a^{2}}\epsilon\zeta(\partial\zeta)^{2}\right.\nonumber\\
  & &\!\!\!\!\left.  - (\epsilon - \eta)\zeta^{2}\partial^{2}\zeta - 2\epsilon\left(1 - \frac{\epsilon}{4}\right)\dot{\zeta}\partial_{i}\zeta\partial_{i}\partial^{-2}\dot{\zeta}\right.\nonumber\\
  & & \!\!\!\!\left. + \frac{\epsilon^{2}}{4}\partial^{2}\zeta\partial_{i}\partial^{-2}\dot{\zeta}\partial_{i}\partial^{-2}\dot{\zeta}\right]\,,
\end{eqnarray}

The numerical calculation of the bispectrum is technically challenging
and is described in more detail in \cite{Horner:2013sea}. Briefly, for the
equilateral configuration it requires the calculation of the
following integral
\begin{equation}\label{fnl_prelim}
f_{\mathrm{NL}} = \frac{1}{3|\zeta|^{4}}\times{\cal I}\left[\zeta^{*3}\int_{N_{0}}^{N_{1}} dN\, (f_{1}\zeta^{3} + f_{2}\zeta\zeta^{\prime 2})\right]\,,
\end{equation}
where $\zeta = \zeta_{k}$, $\zeta^{\prime} =
\mathrm{d}\zeta/\mathrm{d}N$, and $\cal I$ represents the imaginary
part. The background functions $f_{i}$ are given by
\begin{eqnarray}
\!\!\!\!\!f_{1} & = & \frac{5k^{2}a\epsilon}{H}(2\eta - 3\epsilon)\,,\nonumber\\
\!\!\!\!\!f_{2} & = & -5Ha^{3}\epsilon\left(4\eta - \frac{3}{4} \epsilon^{2}\right)\,.
\end{eqnarray}
The times $N_{0}$ and $N_{1}$ correspond to when the mode is
sufficiently sub- and super-horizon respectively. For calculating the
shape dependence we restrict ourselves to the case of isosceles
triangles so we parametrise our modes in the following
way. $|\textbf{k}_{1}| = |\textbf{k}_{2}| = k, |\textbf{k}_{3}| =
\beta k $. This covers most configurations of interest ($\beta = 0$ is
squeezed, $\beta = 1$ is equilateral, $\beta = 2$ is folded) and is
simple to interpret.
\begin{figure*}[t]
  \begin{center}
    \begin{tabular}{cc}
      \makebox[8.5cm][c]{
        \includegraphics[width=8.5cm,trim=0cm 0cm 0cm
        0cm,clip,angle=0]{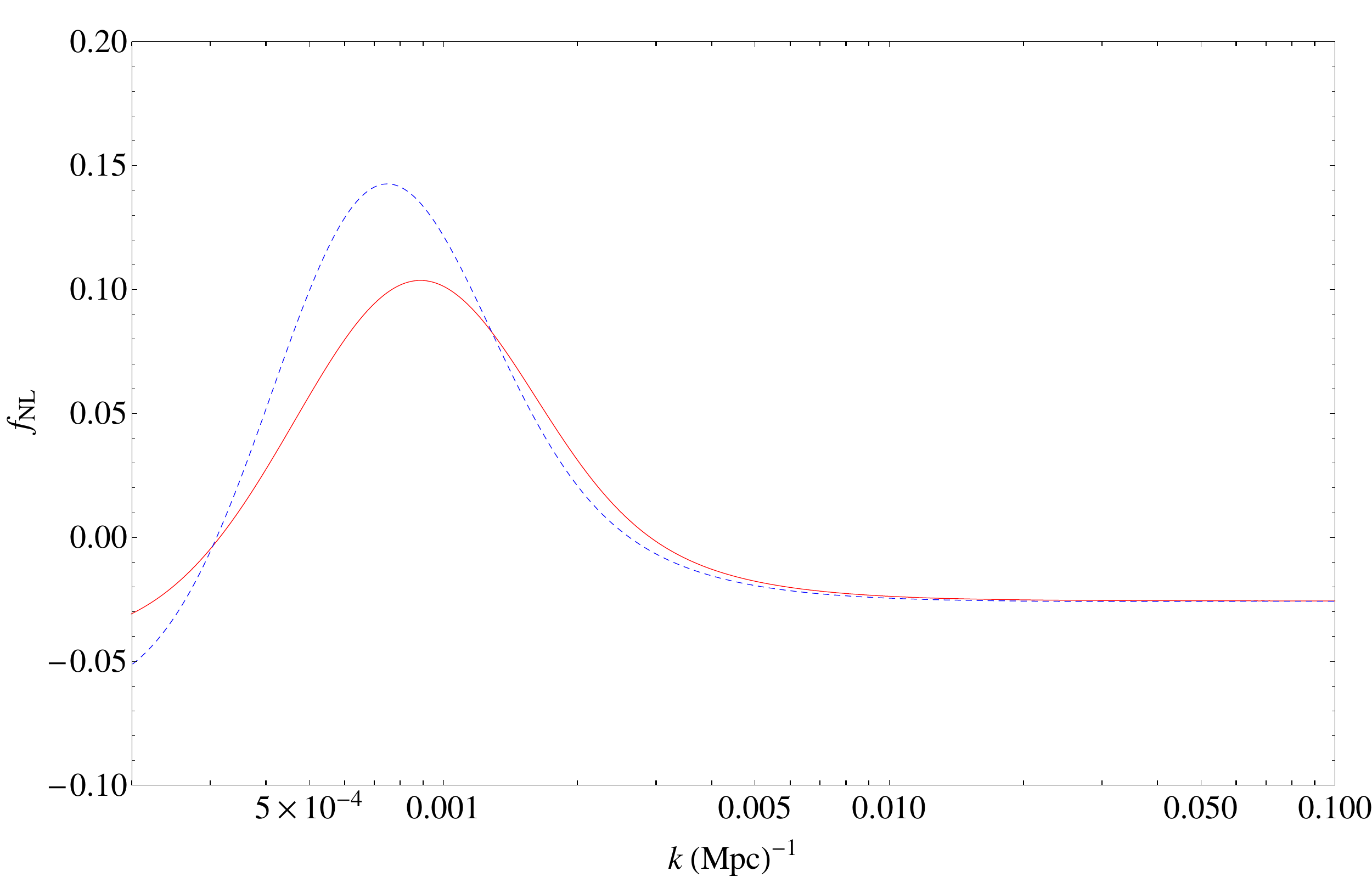}}&
      \makebox[8.5cm][c]{\includegraphics[width=8.5cm,trim=0cm 0cm 0cm
        0cm,clip,angle=0]{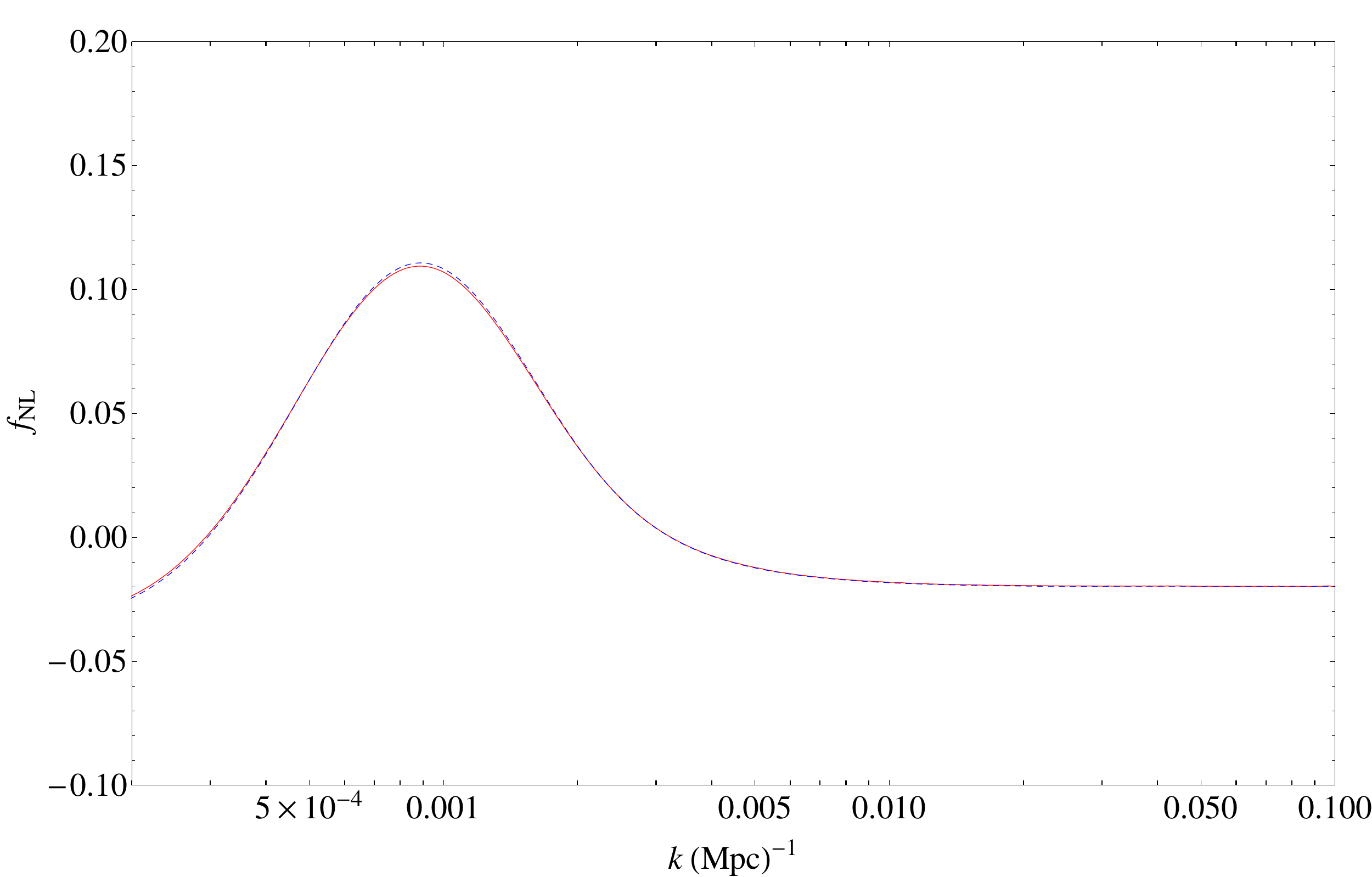}}
    \end{tabular}
    \caption{\fnl as a function of $k$ for equilateral (left) and
      squeezed (right) configurations. The blue (dashed) curves
      represents the numerical calculation. The red curves represent
      the slow roll approximation (\ref{SRapprox}) (left) and the consistency
      condition $5/12 (n_{s} - 1)$ (right). It is not
      possible to calculate the exact squeezed configuration
      numerically so a configuration with  $\beta = 0.1$ was used to
      approximate the squeezed limit. }
    \label{fig:fnl1}
  \end{center}
\end{figure*}

\section{Results}\label{results}

For the toy-model given in (\ref{toy_model}) the
non-Gaussianity amplitude is plotted in Fig.~\ref{fig:fnl1}. For
comparison, as well as a consistency check, we plot the full-numerical
calculation (blue-dashed) as well as the the slow-roll approximation
(red-solid) which, in the equilateral limit, is given by \cite{Maldacena:2002vr}
\begin{equation}\label{SRapprox}
f_{\mathrm{NL}}(k) = \frac{5}{12}\left(n_{s}(k) - 1 + \frac{5}{6}n_{t}(k)\right)\,.
\end{equation} 

\begin{figure}[t]
  \begin{center}
    \includegraphics[width=8.5cm,trim=0cm 0cm 0cm 0cm,clip]{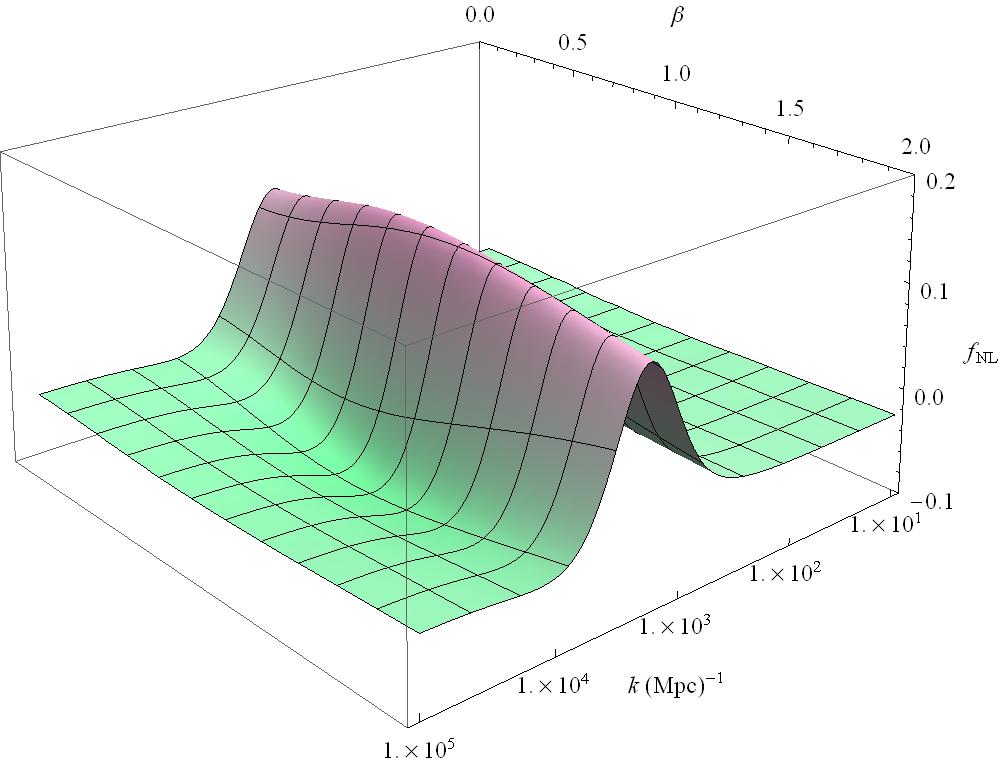} 
    \caption{ \fnl as a function of scale $k$ and shape $\beta$. There
      is a mild peak in the equilateral limit, $\beta = 1$. For all
      shapes the non-Gaussianity peaks around the scales corresponding
      to the size of the horizon at the time when the background acceleration is
      changing.}
    \label{fig:fnl2}
  \end{center}
\end{figure}

In applying this formula we used the exact values of $n_{s}$ and
$n_{t}$ given by equations \ref{nsr_numeric}. As can be seen from
Fig.~\ref{fig:fnl1}, if values close to $r\sim 0.2$ are confirmed from
polarisation measurements, the non-Gaussianity on large scales are
likely to be an order of magnitude larger than expected. This is
simply because $r \propto \epsilon$ but on smaller scales $\epsilon$
is constrained to be lower by the total intensity measurements. The
only way to reconcile the two regimes is by having $\epsilon$ change
to a lower value at later times and this results in an enhancement of
non-Gaussianity being generated as the value is changing.
Fig.~\ref{fig:fnl1} also shows that, even with strong scale
dependence, there is remarkable agreement between the full numerical
results and the Maldacena formula, with deviations only occurring at
the largest scales. Fig.~\ref{fig:fnl2} shows the complete scale and
shape dependence of \fnl.

\section{Discussion}\label{conclusion}

Models of inflation that contain a feature causing the background
acceleration to change can reconcile \planck\ and \bicep\ observations of
the CMB total intensity and polarisation power spectra. We have shown
that these models result in enhanced non-Gaussianity at scales
corresponding to the size of the horizon at the time when the
acceleration is changing. The level of non-Gaussianity at these scales
is an order of magnitude larger than what is expected in the
standard case with no feature and is strongly scale dependent. 

Whilst the effect was illustrated using a simple toy-model of the
background evolution $H(t), \epsilon(t)$, etc, we expect the
non-Gaussian enhancement to be present in any model where the
acceleration changes relatively quickly in order to fit the \planck\ and
\bicep\ combination. The exact form of non-Gaussianity will obviously
be model dependent.

It is not clear that this level of non-Gaussianity will be observable
since it corresponds to scales $\ell \sim 2\to 80$ where there may not be a
sufficient number of CMB modes on the sky to ever constrain \fnl to
${\cal O}(10^{-1})$. However cross-correlation with other surveys of
large scale structure may help to constrain non-Gaussianity on these
scales. In particular it may be possible to detect any anomalous
correlation of modes induced by the non-Gaussianity.

The biggest question at this time however is whether or not the
claimed detection of primordial tensor modes by \bicep\ is
correct. This will be addressed in the near future as the polarisation
signal is observed at more frequencies at the same signal-to-noise
levels reached by the \bicep\ experiment. 

\begin{acknowledgments}
  We thank Marco Peloso for useful discussions. JSH is supported by a
  STFC studentship. CRC and JSH acknowledge the hospitality of the
  Perimeter Institute for Theoretical Physics and the Canadian
  Institute for Theoretical Astrophysics where some of this work was
  carried out.
\end{acknowledgments}

\bibliography{paper}

\end{document}